# Future green technology: freezing water micro-droplet as an all-optical switch based on time domain photonic hook

Oleg V. Minin[1], Yinghui Cao[2] and Igor V. Minin[1],*,

[1]Nondestructive Testing School, Tomsk Polytechnic University, 36 Lenin Avenue, Tomsk 634050, Russia
[2]College of Computer Science and Technology, Jilin University, 2699 Qianjin Street, Changchun 130012, China.

* Correspondence: prof.minin@gmail.com;

**Abstract**
Here we show that mesoscale freezing water droplet with low Bond number can behave as fully biocompatible natural microlenses to form the photonic hook in application to tunable optical switch. We first introduce and demonstrated the basic concepts of all-optical switch without changes in the wavelength of illumination of a particle or any moving parts involved. The principle of operation of the switch based on the temperature-induced phase change inside water droplet refractive index. Simulation shows that the optical isolation of switched channels for an optical switch with linear dimensions of about $15\lambda^3$, based on freezing water droplet, can reach 10 dB during the temperature variation at fixed wavelength. The freezing mesoscale droplets acting as time domain photonic hook generator open the intriguing route for full optical switching in a multifunctional green electronics tool for sensing, integrated optics and optical computers.

**Keywords:** water droplet, freezing, photonic hook, mesotronics,

**Introduction**
The field of optical switching has a long history and is too broad [1]. Several technologies has been studied for spatial or wavelength-selective optical switches [2]. A critical comparison of various types of optical switches was analyzed in [3]. Instead of a wavelength-selective optical switch, it is possible to use switching the spatial orientation of the recently discovered (2015) the photonic hook phenomenon, which made it possible to numerically [4,5] and experimentally [6] demonstrate the concept of a new all-optical switch.
At the same time, such natural material as a water have shown perspectives for use in future green technologies [7-9]. Using a single water droplet as a photonic component with different optical properties has become a new intriguing paradigm in mesotronics (mesoscale photonics) for the wavefront manipulation and unusual applications [10-15].
Phase-change materials as a freezing water droplet are suited to fulfill all optical switching function due to their temperature sensitivity and relatively fast switching dynamics. It's allow to obtain a curved photon flux uses geometrically symmetric (spherical) water droplet with refractive index asymmetry due to freezing. Such freezing droplets are the so-called Janus particles obtained by combining two materials (water and ice in our case) with different optical properties [16].

Recently we demonstrated the possibility of generating a self-bending photonic hook in the time domain (TD-PH) based on a freezing water drop [17]. A key property of a photonic hook for realization the function of an all-optical switch is the dependence of the curvature of its beam vs the position of water-ice interface during the freezing of the droplet. Therefore, with a certain internal dynamically change structure of a freezing droplet and reception zones, it is possible to achieve a change in the magnitude of the optical signal in each of the channels when the water phase changes during the freezing.

**Model**

As in [17] the formation of localized curved beam in the form of TD-PH by freezing mesoscale water droplet immersed in air (n = 1) was studied by using the commercial software Comsol Multiphysics based on the finite elements method. A few words about the shape of the droplet [18]. The Bond number ( Bo ) characterizes the effects of surface tension and gravity:

$$\text{Bo} = \frac{\Delta \rho g R^2}{\sigma},$$

where: $\sigma$ is the surface tension of the droplet, $\Delta\rho$ – the density difference between droplet and surrounding medium, g is a gravitational constant and $R$ is the radius of the droplet.

It could be noted, that the contact angle does not include into the Bond number (which is more than 150 degree for super-hydrophobic surface), so Bo is unable to characterize effects of interaction between solid and liquids. The minimum droplet radius on the surface decreases with the Bond number decrease, but the behavior of the minimum droplet radius with the Bond number is not linear due to the energy balance [19]. In micron-sized water droplets, the Laplacian (capillary) pressure is much greater than the hydrostatic pressure (due to gravity). In this case, we can write:

$$\frac{\sigma}{R} \gg \rho g R \text{, thus } R \ll \sqrt{\frac{\sigma}{\rho g}} = \sqrt{\frac{72 \cdot 10^{-3}}{10^3 \cdot 9,8}} \approx 2,7 \text{ mm}.$$

So for the micron sized droplet the shape of the droplet with small Bond number may be considered as an almost spherical [17,20], which has a minimum surface area for a fixed volume.

The process of droplet solidification and freezing are dependent on the following key factors: physical, dynamic and thermal factors. An experimental study on the dynamics of the water-ice interface evolution upon freezing was discussed, for example, in [21]. In addition, the shape of the water droplet deposited on the cooled surface cannot be an ideal sphere in some cases [21]. It was shown that when the interface moves up above the center, its curvature change and is no longer a spherical curve. Importantly, the experiment of water drop freezing in [21] was performed in a Hele-Shaw cell apparatus, which consists of two parallel flat plates separated by a small gap, in between which the droplet is freeze. Although the experimental studies discussed above were related to millimeter-scale droplets, the experimental conditions corresponded to a two-dimensional model. Therefore, in contrast to [17], we introduced a small distortion of the shape of a spherical droplet and the shape of the water-ice interface according to the results of [21] proportionally to the droplet radius.

As in [17], we use 2D geometry in simulations to reduce the computational time. The Perfect Matched Layer (PML) is applied. Outside and inside the droplet, the mesh size is λ/8 and λ/15, respectively. The incident plane wave propagating along the x-axis was linearly polarized along the y-axis. It was assumed that ice and water are materials without any inhomogeneity's and inclusions. The refractive indices of ice and water at the wavelength of λ = 589 nm are 1.201 and 1.334, respectively [22]. A schematic diagram of the optical switch based on time domain photonic hook by freezing water droplet shown in Figure 1.

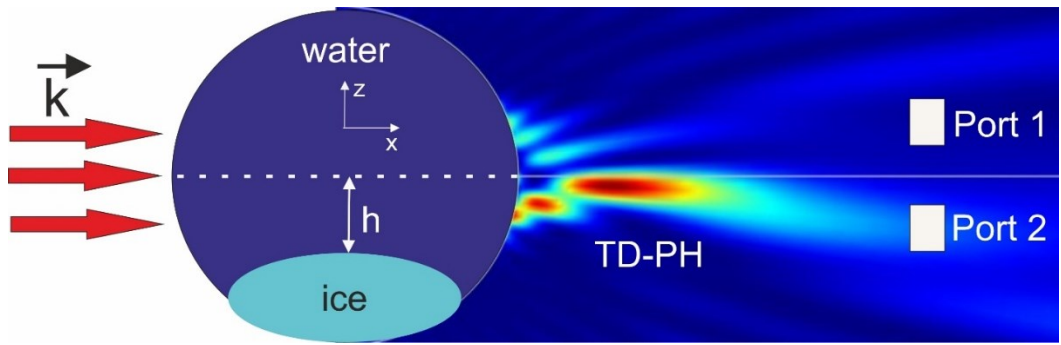

**Figure 1.** A schematic diagram of the time domain photonic hook formation by freezing water droplet for all optical switch. Two-channel receivers Port 1 and Port 2 schematically shown as a white squares; h — position of the ice-water interface.

### Results and discussion

The evolution of the photonic hook shape and water-ice interface (h,$R_{in}$) during freezing of water droplet is shown in Fig. 2 for a drop with a radius of R=2.5 microns. Such scenario of water-ice interface propagation at different times corresponds to the model discussed in [20,21].

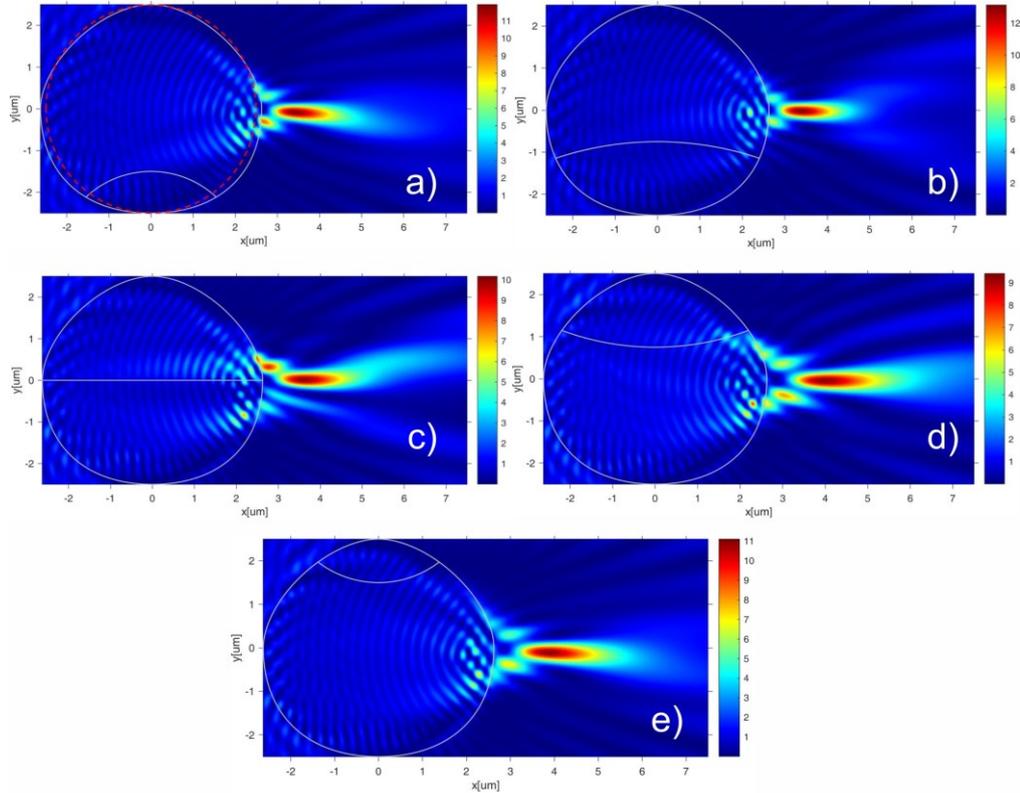

**Figure 2.** The evolution of the photonic hook shape and water-ice interface during freezing of water droplet: red circle show the spherical droplet shape and white counter corresponds to the distortion of the droplet shape with diameters of $D_x$=5.25 um and $D_y$=5 um. White curve inside droplet shown the shape and position h of the water-ice interface. It is shown the electric field intensity enhancement $(E/E_0)^2$. a) h=-1.5 um, $R_{in}$=0.9R, b) h=-0.75 um, $R_{in}$=3R, c) h=0, d) h=0.75 um, $R_{in}$=2.5R, e) h=1.5 um, $R_{in}$=0.9R.

In general, the dynamics of change and formation of the photonic hook coincides with the previously considered properties of a spherical water droplet [17]. The unique properties of the TD PH is on change the angle of its bending and curvature with respect to the temperature of the droplet and maintain an high spatial localization at distances more greater than the Rayleigh diffraction length [23, 24]. One can see (Fig.2) the curvature and the shape of the water-ice interface changes with different h position of the interface. It is also clearly seen that the propagation path of the localized field at the shadow side of the droplet is deflected at the inflection point near the position of $I_{max}$=max($|E/E_0|^2$), resulting in bending of TD PH. Note that the water-ice interface is proportional to a square root of freezing time [20].

Due to the asymmetry of freezing water-ice droplet, the bottom half of the drop contains less ice than the top half at initial stage, thus photonic hook is bent downward to the bottom of the droplet when h = -1.5 um (Fig.2a). For h=0 (Fig.2c), the top half droplet is water and the bottom half droplet is ice, so the top half droplet is optically thicker than the bottom half droplet due to differences in refractive index, and the photonic hook is bent upward. For large values of h, the situation with the orientation of the photonic hook is reversed and the TD PH is bent down again (Fig.2e).

Such dynamics of changing the orientation of the photonic hook during water droplet freezing makes it possible to use the considered effect to create an optical switch.

The electric field intensity distributions along y-axis and relative energy transmission S for two-channel receivers, calculated as S = 10*log($W_1$/$W_2$) at Port1 (0.1um<y<0.7um) and Port2 (-0.7um<y<0.1um) (see Fig.1) for different position of Port1 and Port2 plane along X-axis are shown in Figure 3.

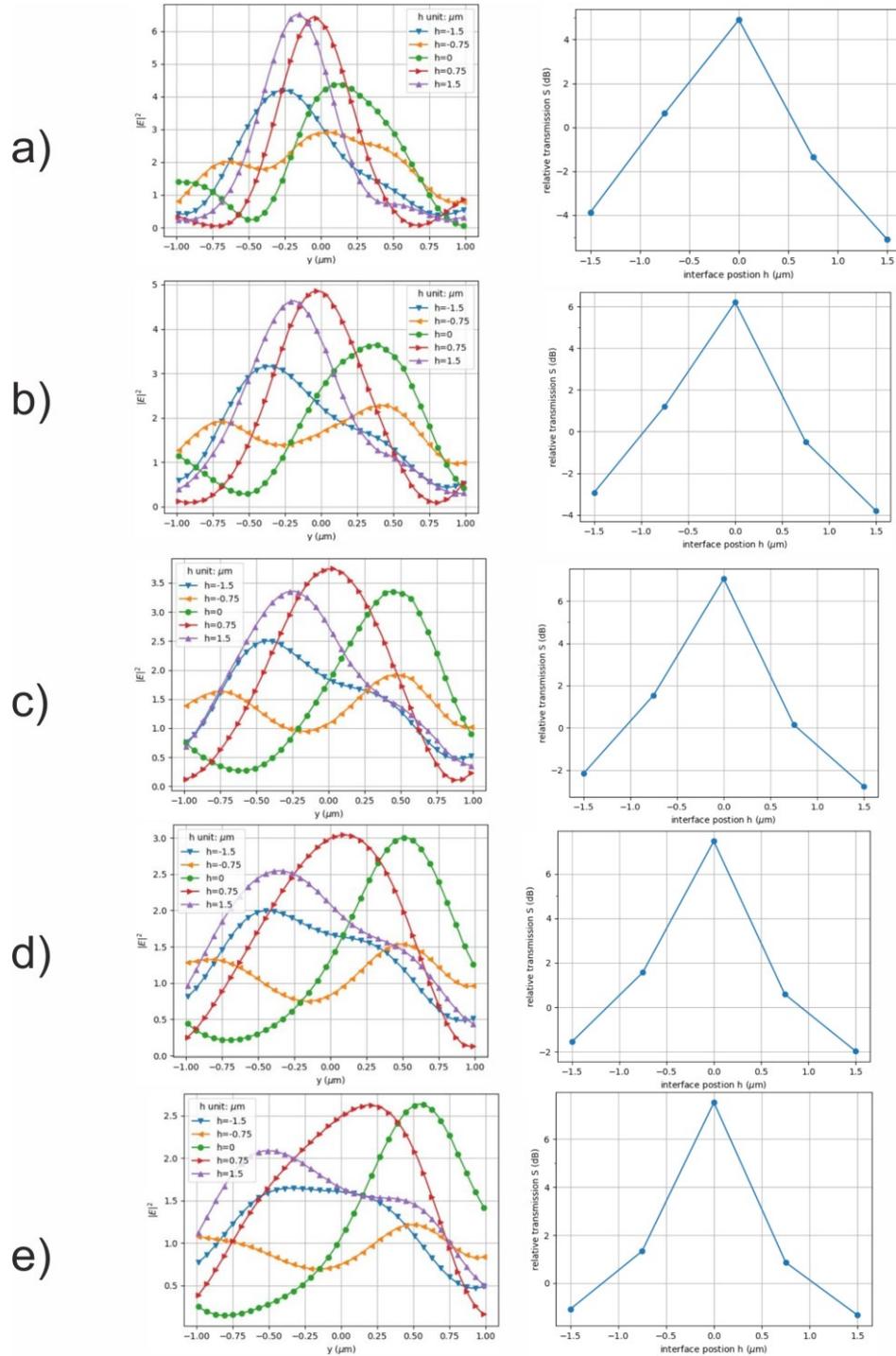

**Figure 3.** Electric field intensity distributions along y-axis (left column) and relative transmission S (right column) for different position of Port 1 and Port 2 line along X-axis: a) x= 5.0 um, b) x=5.5 um, c) x=6.0 um, d) x=6.5 um, e) x=7.0 um.

Following to [4], here is the dependence of the relative energy transmission $S=\log(W_1/W_2)$ on the water-ice interface position inside droplet, where the energies $W_{1,2}$ are the integral of the electric field intensity over the cross-sectional area $\Sigma_{1,2}$ of the corresponding receiving port (1 or 2):

$$W_{1,2} = \int_{\Sigma_{1,2}} |E|^2 \, d\sigma.$$

The extremes on these curves corresponds to the alternative states of the optical switch, when the signal $S_2$ from the second Port 2 (see Fig. 1) or the signal $S_1$ from the first Port 1 prevails. In Fig. 1, such states are marked with large white squares.

One can see from Fig.3 that when the position of the water-ice interface boundary (h) changes, the field intensity maximum in the plane of Ports 1 and 2 changes its position. This trend is the same for different observation planes along the x-axis, but the law of change in the position of the field intensity maximum is somewhat different. Accordingly, the maximum value of relative energy transmission also changes.

This statement is confirmed by Fig. 4, which shows the position of the field intensity maximum in the plane of Ports 1-2 for various distances from the shadow surface of the droplet (a) and the dependence of the maximum of the relative energy transmission S and the maximum field intensity vs the position of the observation plane (b). It is quite obvious that when the observation plane (the position of the Ports along x-axis) moves away from the shadow surface of the drop, the value of the maximum relative energy transmission increases with a decrease in the maximum value of the field intensity enhancement. At the same time, the signal difference $dS = S_1 - S_2$ may be serve as a measure of the optical decoupling (isolation) of switching channels [4]. One can see (Fig.5) that the maximum relative energy transmission $dS$ can reach a value of 10 dB under the considered conditions. We emphasize that we are not optimizing the optical switch, but only demonstrating a possible physical concept. Larger S values can be achieved by using other droplet diameter or materials when the refractive index contrast is higher than that considered.

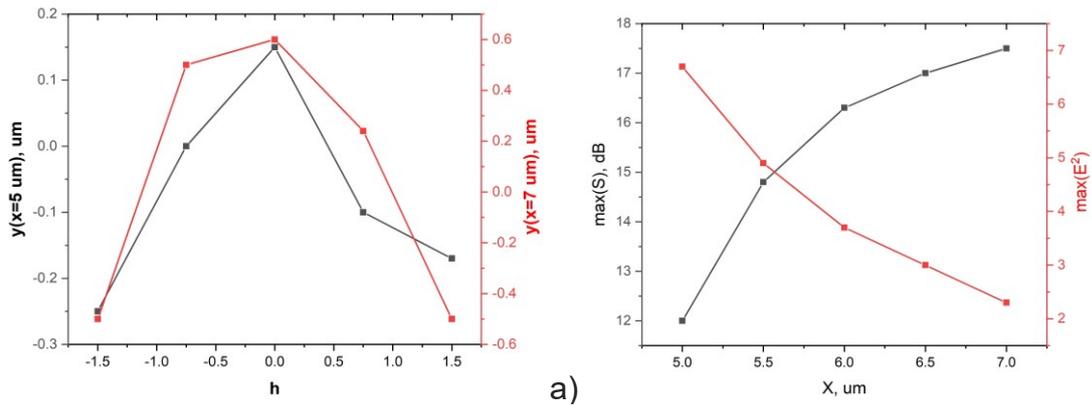

**Figure 4**. The position of maximal field intensity along y-axis for x=5 um and x=7 um vs water-ice interface position h (a); Maximal of the relative energy transmission $S$ and maximal field intensity enhancement vs position of the Ports plane along x-axis (b).

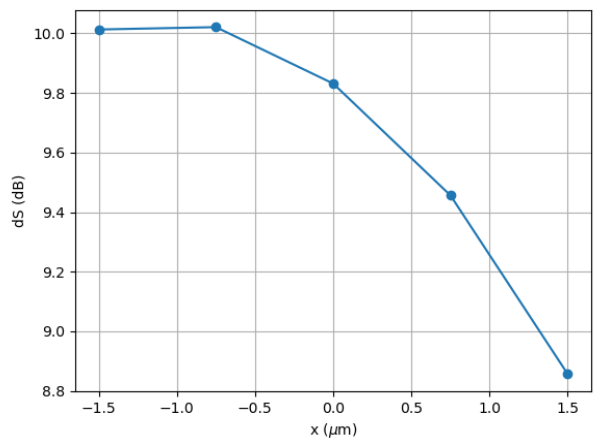

**Figure 5**. Relative energy transmission $dS = S_{h=0} - S_{h=1.5um}$ of receiving ports versus the position of the Ports plane along x-axis.

As for the operating speed of the optical switch, a one-dimensional analytical solution of the freezing of a millimeter-scale spherical water droplet moving in cold air was discussed in [25] in the case of a uniform temperature distribution over the liquid's core [26]. A 3-dimensional simulations shown the fast freezing dynamics of micro-sized water droplets with radiuses of 5 μm and 0.5 μm [27]. With a decrease in the diameter of a water drop, multiple nucleation centers are formed, which accelerate the freezing of the drop. For 50 μm droplet the freezing time is in order of 1 ms [28]. This time is comparable with liquid crystal based optical switch [29].

**Conclusion**

We have successfully demonstrated a new method of generating mesoscale freezing droplets that can work as flexible and temperature tunable all-optical switch. Such a switch can be attributed to the class of thermo-optic type [30-31]. The phenomenon of TD PH is key to the efficient manipulation of light at mesoscale based on natural materials. We have shown the fundamental possibility of developing an all-optical mesoscale (with a size of the order of the wavelength) two-channel switch based on freezing water droplet. Due to the unique property of the time-domain photonic hook to change the curvature with respect to the temperature, this switch is a perspective candidate for the implementation of optical switching in future miniature 'on-a-chip' devices and green optoelectronics and mesotronics [32] without the use of any micromechanical devices and in the absence of electrical signal control or multi-wavelength illumination.

It could be noted that the larger S values can be achieved by taking onto account that in freezing droplet, the ice may be a porous medium, rather than a complete solid ice due to the bubble formation inside the droplet [33], and the refractive index contrast may be higher than that considered. Moreover, since a water as a phase change material are fluid in the liquid state, it is advisable to use microencapsulation methods to ensure the reusability of the switch. Microencapsulation makes it possible to obtain a material with a phase change, enclosed in capsules with sizes less than 1 micron [34,35].

It should be noted that in this work, only a proof-of-concept for all-optical miniature water based time domain photonic hook switching is presented, and we do not carry out a full-scale optimization of the full optical switch characteristics. The authors hope that a future all-optical computer, which uses "green" mesoscale photonics instead of electron, comes a step closer.

**Author Contributions:** Conceptualization, O.V.M. and I.V.M.; methodology, O.V.M. and I.V.M.; software, Y.C.; formal analysis, O.V.M. and I.V.M.; investigation, O.V.M., Y.C. and I.V.M.; writing—original draft preparation, O.V.M. and I.V.M.; writing—review and editing, O.V.M. and I.V.M.; visualization, I.V.M, O.V.M., Y.C.; supervision, I.V.M. and O.V.M.; All authors have read and agreed to the published version of the manuscript.
**Institutional Review Board Statement:** Not applicable.
**Informed Consent Statement:** Not applicable.
**Data Availability Statement:** The data that support the findings of this study are available from the corresponding author upon reasonable request.
**Acknowledgments:** I.V.M. and O.V.M. acknowledge the Tomsk Polytechnic University Development Program.
**Conflicts of Interest:** The authors declare no conflict of interest.